\documentclass[prb,aps,twocolumn,showpacs,superscriptaddress,amsfonts,amsmath,fleqn,floatfix]{revtex4-1}

\usepackage{graphicx}
\usepackage{dcolumn}
\usepackage{bm}
\usepackage{float}

\usepackage[breaklinks]{hyperref}
\usepackage[all]{hypcap}

\hypersetup{
    linktoc=section,
    colorlinks=true,       
    linkcolor=blue,          
    citecolor=blue,        
    filecolor=magenta,      
    urlcolor=blue           
            }

\begin{document}
\title{{Gap like structure in a nonsuperconducting layered oxycarbonate\\Bi$_{2+x}$Sr$_{4-x}$Cu$_2$CO$_3$O$_{8+\delta}$ single crystal}}

\author{S. I. Vedeneev}
\affiliation{Laboratoire National des Champs Magn\'etiques Intenses, 25 avenue des Martyrs, CNRS-UJF-INSA-UPS, 38042
Grenoble, France} \affiliation{P. N. Lebedev Physical Institute, Russian Academy of Sciences, 119991 Moscow, Russia}
\author{B. A. Piot}
\affiliation{Laboratoire National des Champs Magn\'etiques Intenses, 25 avenue des Martyrs, CNRS-UJF-INSA-UPS, 38042
Grenoble, France}
\author{D. K. Maude}
\affiliation{Laboratoire National des Champs Magn\'etiques Intenses, 25 avenue des Martyrs, CNRS-UJF-INSA-UPS, 38042
Grenoble, France}

\date{\today}

\begin{abstract}
The magnetic field and temperature dependence of the in-plane
tunneling conductance $dI/dV(V)$ in high-quality
nonsuperconducting (down to $10$~mK) layered oxycarbonate
Bi$_{2+x}$Sr$_{4-x}$Cu$_2$CO$_3$O$_{8+\delta}$ single crystals has
been investigated using break junctions. Combining measurements of
the in-plane magnetoresistivity $\rho_{ab}(T,H)$ and the
magnetotunneling, we present evidence for the existence of a small
"pseudogap" in a nonsuperconducting cuprate, without local
incoherent pairs or any correlation phenomena associated with
superconductivity. We are unable to distinguish if such a
"pseudogap" is totally unrelated to superconductivity or if its
existence is a necessary condition for the subsequent occurrence
of superconductivity with increasing carrier density in the
sample.
\end{abstract}

\pacs{74.72.-h, 74.50.+r}

\maketitle
\section{Introduction}

It is currently well established that in the normal state of underdoped high-$T_c$ superconductors (HTS) there is a
pseudogap state, in which the density of states is depleted upon decreasing temperature below a characteristic
temperature $T^{*} \gg T_c$. Many experiments (for a review, see e.g. Ref. [\onlinecite{Basov05}]) have investigated
this problem, however, the data and their interpretations are still controversial. This is the all-important question
of high temperature superconductivity since it connects normal state correlations (referred to the pseudogap) to the
origin of the high $T_c$. At present, there are two markedly different, but nevertheless plausible, explanations of the
pseudogap in the normal state. One is associated with the formation of local incoherent pairs above $T_c$, that is, the
pseudogap is a precursor to superconductivity (see \emph{e.g.} Refs. [\onlinecite{Nakayama09,Fischer07}]), and in the
other the pseudogap competes with superconductivity (see \emph{e.g.} Ref. [\onlinecite{Kondo09}]). Whether the
pseudogap state is the precursor to superconductivity or a state that competes with superconductivity has been a matter
of long-standing debates.

Very recently Dubroka \textit{et al.}\cite{Dubroka11} based on studies of the infrared c-axis conductivity of
\textit{R}Ba$_2$Cu$_3$O$_{7-\delta}$ (\textit{R} = Y, Gd, Eu) have suggested that the anomalous normal-state properties
of underdoped cuprate superconductors are determined by two distinct correlation phenomena. One of them is due to the
competing pseudogap below temperature $T^* \gg T_c$, whereas the other exhibits a signature of the precursor
superconducting state at $T_c < T < T^{ons}$. In this state a growth of the spectral weight of the coherent response
was observed. The magnetic field had a similar effect on the coherence of an electronic state at $T_c < T < T^{ons}$ as
it has at $T < T_c$. These results are consistent with other work~\cite{Kondo11} which also demonstrated that in
Bi$_2$Sr$_2$CaCu$_2$O$_{8+\delta}$ (Bi2212) and Bi$_{2+x}$Sr$_{2-x}$CuO$_{6+\delta}$ (Bi2201) in the pseudogap region
two different states coexist; one is due to pair formation and persists to an intermediate temperature $T_{pair} < T^*$
and the second - the ``proper'' pseudogap - characterized by a loss of the spectral weight and anomalies in the
transport properties that extends up to $T^*$.

It is important to note that Dubroka \textit{et al.}\cite{Dubroka11} also surmised that $T^{ons}$ remains finite even
in heavily underdoped samples where $T_c = 0$. This suggestion is in reasonable agreement with our results for the
underdoped Bi2201 cuprates reported in Ref.~[\onlinecite{Vedeneev05}]. By measuring angular dependencies of the
in-plane and out-of-plane magnetoresistivity at temperatures from $1$~K down to $30$~mK, we have obtained evidence for
the presence of vortex like excitations in a nonsuperconducting cuprate in the pseudogap region (in a conventional
superconductor the transition out of the superconducting state is caused by the proliferation of vortices, which
destroy the long-range phase coherence).

On the other hand, as pointed out in Ref.[\onlinecite{Emery95}], HTS are characterized by a relatively small phase
stiffness and poor screening, both of which imply a significantly larger role for phase fluctuations. As a consequence,
in these materials, the transition to the superconducting state may not display a typical mean-field behavior, and
phase fluctuations may have a significant influence on the low-temperature properties. In particular, the onset of
long-range phase order controls the value of $T_c$. In such a scenario, the long-range coherence is lost at $T_c$ due
to fluctuations of the phase of the superconducting order parameter.

Okada  \textit{et al.}\cite{Okada10} have studied the onset temperature of superconducting fluctuations $T_{onset}$ of
Bi$_2$Sr$_{2-x}$\textit{R}${_x}$CuO$_{y}$ (\textit{R} = La and Eu) by measuring the Nernst effect. They have suggested
that the three distinct temperatures, $T_{onset}$, $T^*$, and $T_c$,  are the consequence of a coexisting competing
pseudogap state and incoherent superconductivity, which has been observed below $T_{onset}$. The latter is
qualitatively different from the pseudogap phenomenon that is characterized by $T^*$. The experimentally obtained phase
diagram indicates that the pseudogap state suppresses $T_c$ and enhances superconducting fluctuations.

Recently Grbi\'c \textit{et al.}\cite{Grbic11} have determined the temperature range of superconducting fluctuations
above $T_c$  in YBa$_2$Cu$_3$O$_{7-\delta}$. Their study shows that the temperature range of the superconducting
fluctuations may be as large as $23$~K in a deeply underdoped sample. This relatively wide temperature range for
superconducting fluctuations is qualitatively consistent with the theoretical predictions for phase fluctuations in
underdoped samples.~\cite{Emery95}

On the basis of present knowledge, it is still not well understood how the phenomena characterized by these three
temperatures $T_c$, $T^{ons}$ ($T_{onset}$) and $T^*$ are related to each other and there is currently no consensus
concerning the role of the superconducting fluctuations in the temperature region $T_{onset}
> T_c$.  Hence, the origin of the pseudogap remains unclear but must be understood before HTS problem can be solved.
It is however, experimentally well established that the temperature and magnetic field dependence of the resistivity in
HTS are essentially determined by the normal-state pseudogap.~\cite{Ando02,Ono04,Vedeneev04,Vedeneev07}

In this paper, we attempt to find the pseudogap in a nonsuperconducting cuprate without superconducting fluctuations or
local incoherent pairs and to study the effect of the magnetic field and temperature on this gap using the tunneling
method. This work is motivated by our previous investigation~\cite{Vedeneev05} of vortex like excitations in a non
superconducting Bi2201 single crystals. Measurements were performed on layered oxycarbonate
Bi$_{2+x}$Sr$_{4-x}$Cu$_2$CO$_3$O$_{8+\delta}$ single crystals.

\section{Experiment}

Ceramic oxycarbonate samples with $T_c$ of 30 K were synthesized by intergrowing low-temperature superconductor Bi2201
and nonsuperconducting compound Sr$_4$Cu$_2$CO$_3$. \cite{Pelloquin93} Single crystals of this oxycarbonate were
obtained for the first time in Ref.[\onlinecite{Gorina10}] The Bi$_{2+x}$Sr$_{4-x}$Cu$_2$CO$_3$O$_{8+\delta}$ crystals
used for present study were grown in gas cavities in a KCl solution-melt free growth method. The results of a complex
analysis of the elemental composition, structure, and superconducting properties have been detailed in a recent
publication. \cite{Gorina11} Crystals were mirror-smooth faceted plates with the dimensions $1.5~mm \times (0.4-0.8)~mm
\times (4-6)~\mu$m with the most developed $ab$ planes.

The actual cationic composition of the used samples was measured at 50-70 different points on the crystal and the
scatter in the data was less than 2\%. The X-ray diffraction analysis confirmed the presence of a single phase in our
single crystals. The half-width of the X-ray rocking curves for the crystals was less than $0.1^{\circ}$ and does not
exceed the instrumental resolution.\cite{Gorina11} These data clearly demonstrate the high structural quality and high
homogeneity of the samples on a microscopic scale. We were able to make high quality single-phase superconducting and
nonsuperconducting oxycarbonate single crystals by varying the Bi content and consequently the doping
level.\cite{Vedeneev04} Here we have used the fact that the substitution of trivalent Bi for divalent Sr in the Bi
compounds reduces the hole concentration in the CuO planes.\cite{Harris97}

The $T_c$ value of the
crystals formed by our free growth method can be as high as $36$ K.~\cite{Gorina11} A microprobe analysis of the
investigated samples showed that the nonsuperconducting oxycarbonate single crystals grow at an excess of Bi and have
the composition, normalized according to the formula disregarding the CO$_3$ group, Bi:Sr:Cu:O = 2:4:2:8, 2.45 : 3.56 :
1.38 : 8.61 with the lattice parameter $c=39.289~\AA$.

A four-probe contact configuration, with symmetrical positions of the low-resistance contacts on both $ab$-surfaces of
the samples was used for the measurements of $R_{ab}$ resistance and the tunneling studies. The temperature and
magnetic field dependence of the resistance $R_{ab}(T,H)$ was measured using a lock-in amplifier. For the low
temperature magnetotransport measurements, the crystal was placed directly inside the mixing chamber of a top-loading
dilution fridge. For the in-plane transport current $\mathbf{J}$, a configuration with $\mathbf{H\perp J}$ was
used.~\cite{Vedeneev05}

The sample for the tunneling measurements is fixed on a flexible substrate. In liquid helium, using a differential
screw mechanism, the flexible substrate is bent so that the single crystal breaks along a previously made incision,
resulting in a symmetric tunnel break junction
Bi$_{2+x}$Sr$_{4-x}$Cu$_2$CO$_3$O$_{8+\delta}$-insulator-Bi$_{2+x}$Sr$_{4-x}$Cu$_2$CO$_3$O$_{8+\delta}$. The details of
our break junction setup are described elsewhere~\cite{Vedeneev05a,Vedeneev10}. Mechanically retuning the break
junction repeatedly, we were able to fabricate a large number of tunnel junctions at different places along the initial
break of the crystal where tunneling occurs in the $ab$-plane. A RuO$_2$ thermometer was used to measure the local
temperature of the sample and the temperature was continuously recorded during each measurement.

\section{Results and Discussion}
\subsection {In-plane resistivity $\rho _{ab}$}

\begin{figure}
\includegraphics[width=8.5cm,angle=0,clip]{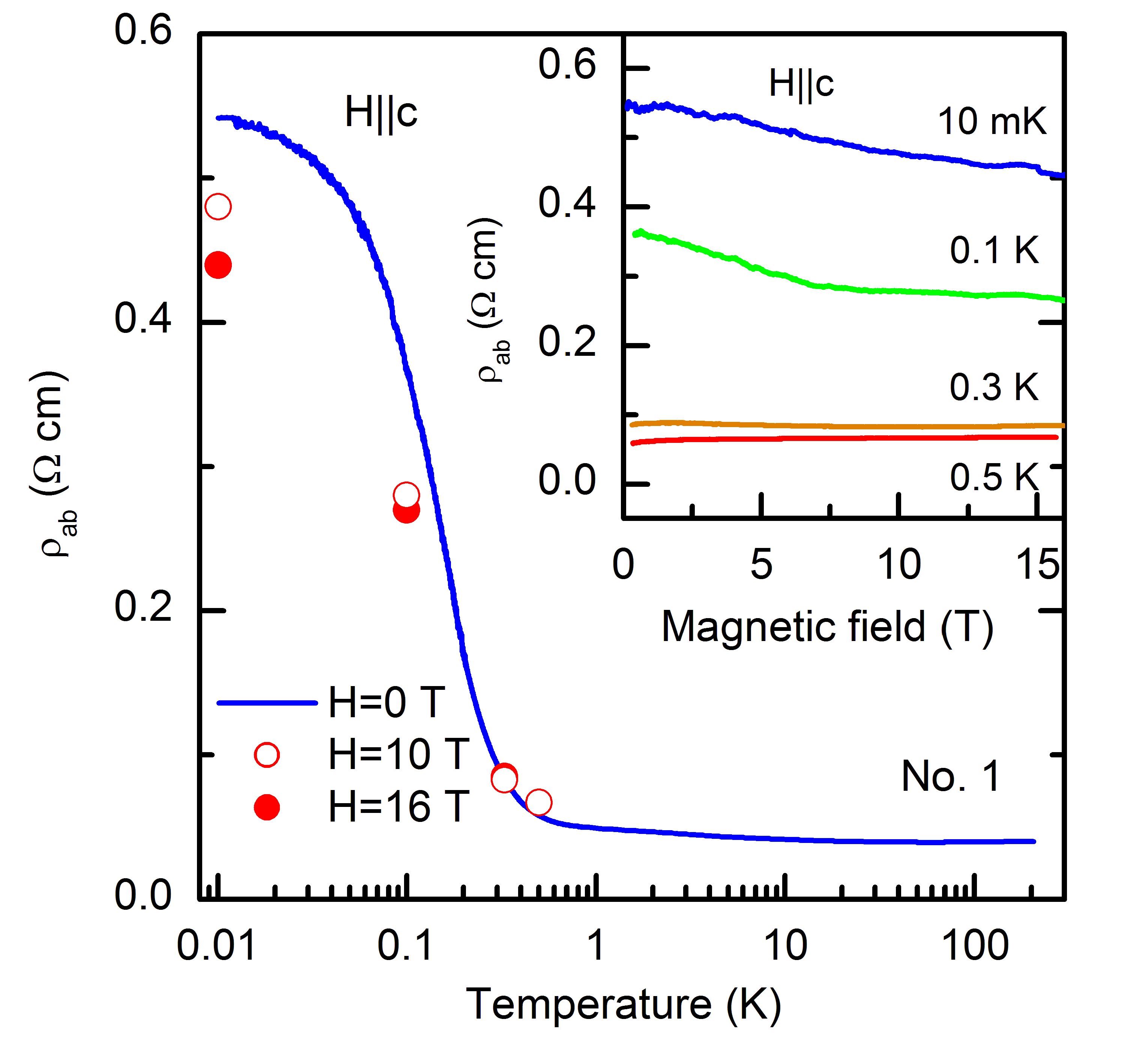}
\caption{\label{fig1} (color online) Temperature dependence of the in-plane $\rho_{ab}$ resistivity for the
oxycarbonate single crystal, with a logarithmic scale for the temperature axis. The data points show the resistivity
data for various values of the magnetic field applied parallel to the $c$-axis. The inset plots the transverse in-plane
magnetoresistance $\rho_{ab}(H)$ for the same single crystal at various temperatures from $10$~mK to $0.5$~K with the
magnetic field parallel to the $c$-axis. (Sample No. 1)}
\end{figure}

Figure~\ref{fig1} (main panel) displays the temperature dependence of the in-plane $\rho_{ab}$ resistivity for the
oxycarbonate single crystal, with a logarithmic scale for the horizontal axis in order to emphasize the low temperature
behavior. The data points show the resistivity data at $H=10$~T and $H=16$~T (open and closed circles, respectively)
with the magnetic field parallel to the $c$-axis. This figure clearly demonstrates that at zero magnetic field, the
sample shows an insulating behavior and remains in its normal state down to 10 mK. At ultra low temperatures, $T=10 -
70$ mK, $\rho_{ab}$ shows a downward deviation (saturation) from the insulating behavior. A similar deviation from the
$\log(1/T)$ dependence of the in-plane resistance of Bi2201 in normal state at ultra low temperatures has been studied
in detail in Ref.[\onlinecite{Vedeneev04}].

The inset in Fig.~\ref{fig1} plots the transverse in-plane magnetoresistance $\rho_{ab}(H)$ for the same single crystal
at various temperatures from $10$ mK to $0.5$ K with the magnetic field parallel to the $c$-axis. At ultra low
temperatures, the magnetic field dependence shows a small ($\simeq 17$~\% at $10$~mK and $16$~T) negative
magnetoresistance and becomes positive with increasing temperature above $0.3$~K (near $\simeq 13$~\% at $0.5$~K and
16~T). The observed magnetoresistance behavior of the oxycarbonate single crystal is in complete agreement with that of
the negative transverse in-plane magnetoresistance reported in investigations of nonsuperconducting Bi2201 single
crystals.~\cite{Jing91} The authors explained these results by localization and the gradual suppression of localization
effects by the magnetic field.~\cite{Lee85} Since our results are in agreement with these experiments, it is reasonable
to assume that they have the same physical origin.

\begin{figure}
\includegraphics[width=8.5cm,angle=0,clip]{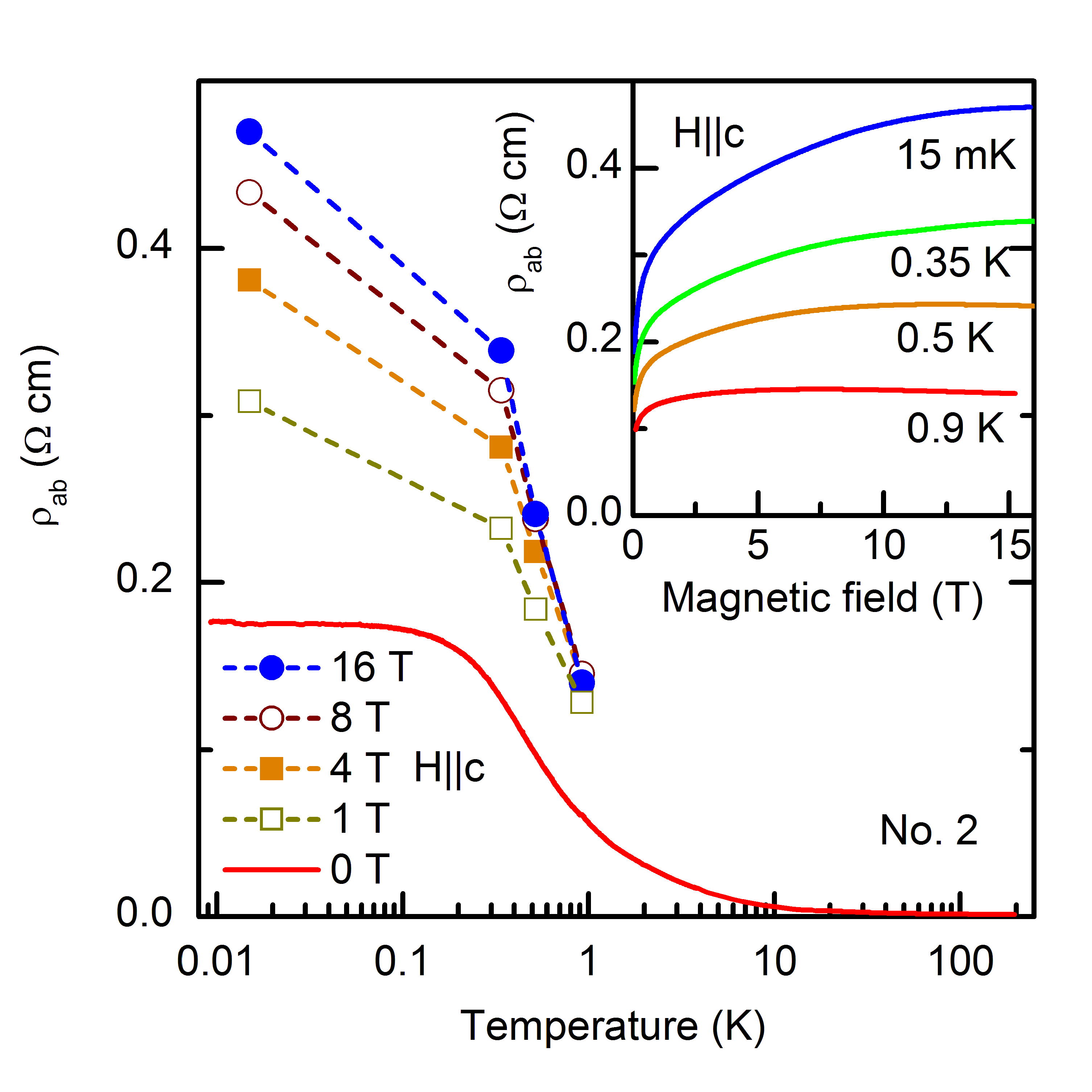}
\caption{\label{fig2} (color online) Temperature dependence of the in-plane $\rho_{ab}$ resistivity for the
oxycarbonate single crystal, with a logarithmic scale for the temperature axis. The data points show the resistivity
data measured at various magnetic fields applied parallel to the c-axis. In the inset the transverse in-plane
magnetoresistance $\rho_{ab}(H)$ is plotted for the same single crystal at various temperatures from $15$~mK to $0.5$~K
with the magnetic field parallel to the $c$-axis. (Sample No. 2)}
\end{figure}

For comparison in Fig.~\ref{fig2}, we show the same data for $\rho_{ab}$ versus the temperature (main panel) and
magnetic field (inset) as in Fig.~\ref{fig1}, but for our oxycarbonate single crystal No. 2 with less excess Bi. As can
be seen from this figure, the sample at zero magnetic field also shows an insulating behavior with a saturation at
ultra low temperatures. However, in contrast to the sample No. 1, the magnetic-field dependence of $\rho_{ab}$ of the
sample No. 2 shows only a positive magnetoresistance, $\simeq 60$~\% at 15~mK and 16~T, (see inset Fig.~\ref{fig2}) in
the temperature region $15$~mK - $0.5$~K. As noted above, in previous studies of Bi2201 single crystals in magnetic
field,~\cite{Vedeneev05} we have shown that the large positive in-plane magnetoresistance of the nonsuperconducting
cuprates in the insulating state is associated with the presence of vortex like excitations.  Similar vortex like
excitations have been observed in superconducting cuprates above the zero-field $T_c$ in magnetic fields by the
detection of a Nernst signal (see \emph{e.g.} Ref.~[\onlinecite{Wang06}]). It is not difficult to see that the sample
No. 2 is unsuitable for clarification of the major issue of high temperature superconductivity; whether the pseudogap
state is the precursor to superconductivity. This is because the temperature $T_{ons}$ ($T^{onset}$ or $T_{pair}$),
below which the incoherent superconductivity state exists (precursor of the superconducting state in which fluctuations
of the phase of the superconducting order parameter or the vortex like excitations can take place), can remain finite
even in heavily underdoped samples where $T_c = 0$. \cite{Dubroka11} All these phenomena prevent an investigation of
the proper pseudogap since we cannot rule out their influence.

The sample No. 1, on the other hand, shows the insulating behavior without any correlation phenomena associated with
superconductivity (Fig.~\ref{fig1}). At $T > 0.3$~K, the magnetic field has practically no effect on $\rho_{ab}$.

\subsection{Gaplike structure}

\begin{figure}
\includegraphics[width=8.5cm,angle=0,clip]{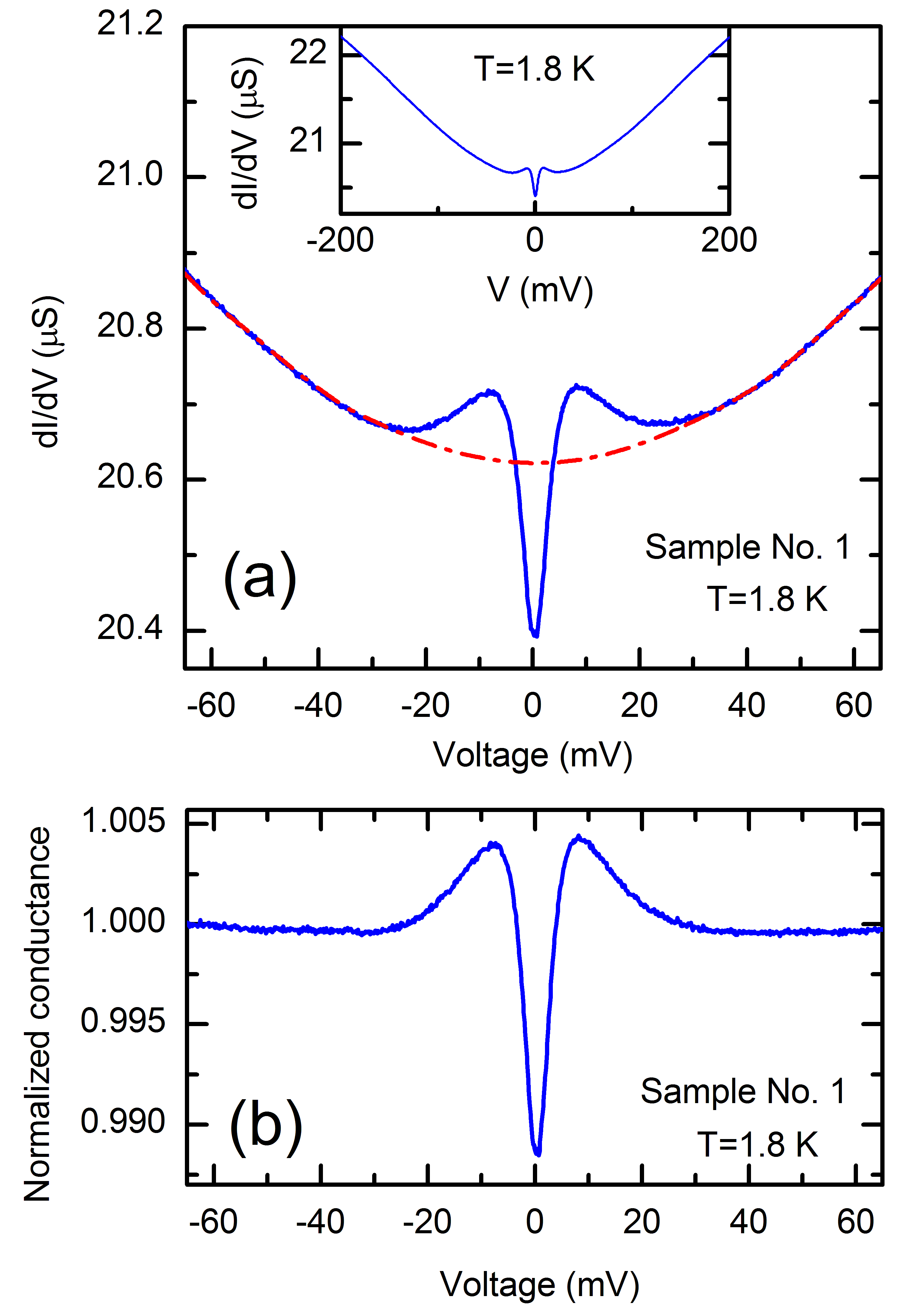}
\caption{\label{fig3} (color online) (a) Tunneling conductance
$dI/dV$ as a function of the bias voltage $V$ for a tunnel break
junction at $T=1.8$~K measured in zero magnetic field (solid
line). The dashed line is a smoothed curve obtained by averaging
the experimental $dI/dV(V)$ curve (see text). The inset shows the
zero-field tunneling conductance at $T=1.8$~K over a larger
voltage range. (b) Tunneling conductance normalized by a smoothed
curve (dashed line in (a)). (Sample No. 1)}
\end{figure}

Figure~\ref{fig3}(a) (main panel) displays the tunneling conductances $dI/dV$ as a function of the bias voltage $V$ for
a tunnel break junction fabricated on Bi$_{2+x}$Sr$_{4-x}$Cu$_2$CO$_3$O$_{8+\delta}$ single crystal (sample No.~1) at
$T=1.8$~K measured in zero magnetic field (solid line). The inset in Fig.~\ref{fig3} shows the zero-field tunneling
conductance at $T=1.8$~K at higher voltage. It can be seen that the conductance is a parabola as required for a
normal-state tunnel structure.~\cite{Wolf85} Note that all the $dI/dV(V)$ investigated here correspond to
$T$-independent resistances at large bias as expected for junctions with a good tunnel barrier without current leakage.
As can be seen in Fig.~\ref{fig3}, $dI/dV(V)$ curves have the shape typical for SIS tunnel break junctions fabricated
on HTS single crystals at $T > T_c$. They have a weak (the conductance does not go to zero) but nevertheless
well-marked gap like structure.

\begin{figure}
\includegraphics[width=8.5cm,angle=0,clip]{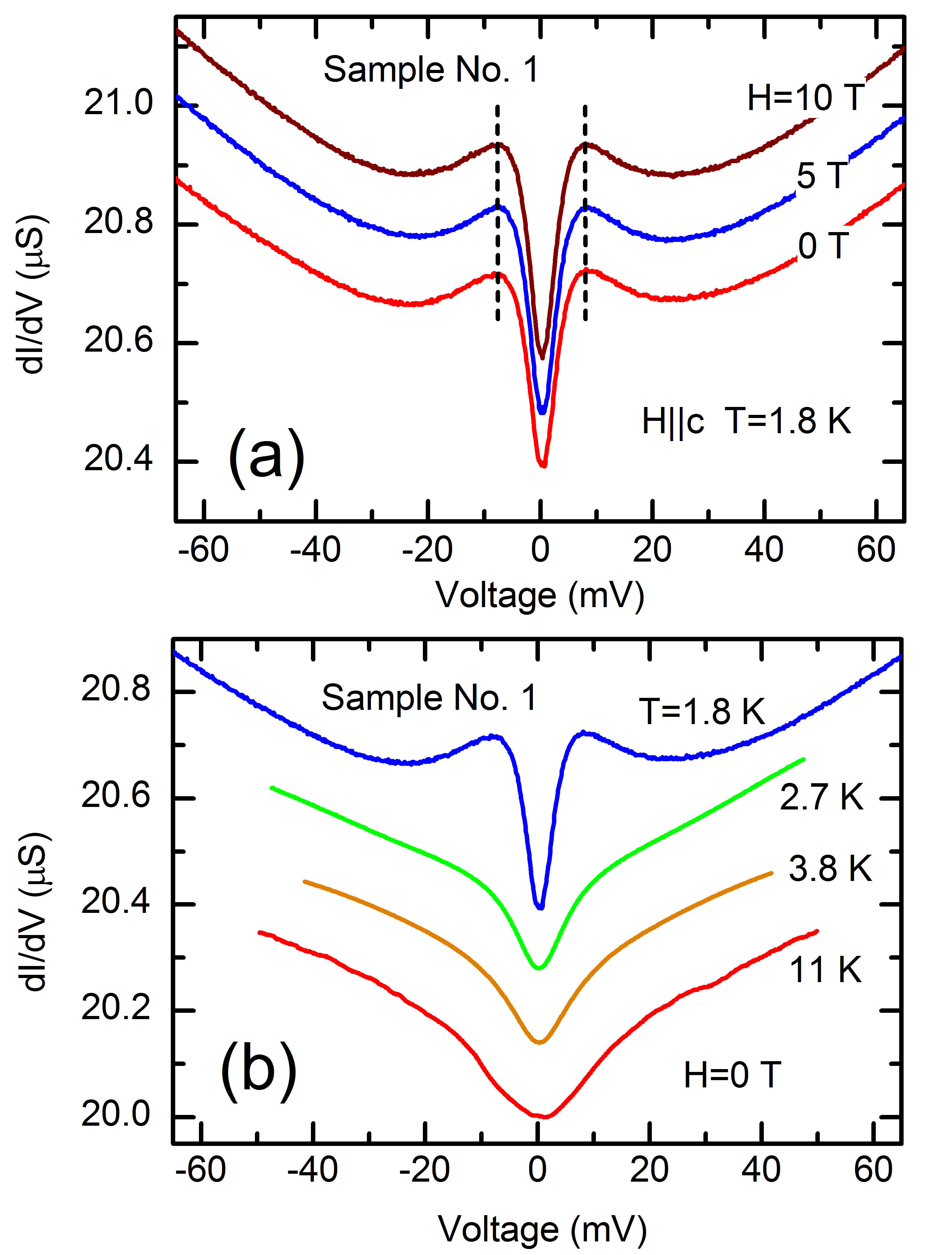}
\caption{\label{fig4} (color online) (a) Tunneling conductances $dI/dV(V)$ for the same tunnel break junction at
$T=1.8$~K for various magnetic fields applied along the c-axis of the crystal. For clarity, the curves have been
shifted vertically relative the bottom one. In magnetic field the gap like structure does not change either in shape or
in the voltage position (dashed vertical line). (b) Tunneling conductances $dI/dV(V)$ for the same tunnel break
junction in zero magnetic field at various temperatures. The curves have been shifted with respect to the upper one.
(Sample No. 1)}
\end{figure}

Since the tunnel conductance in Fig.~\ref{fig3}(a) increases sharply with increasing voltage, we show in
Fig.~\ref{fig3}(b) the normalized tunneling conductance where the gap like structure is better seen. The measured
$dI/dV(V)$ curve has been normalized by a smoothed (averaged) curve (dashed line in Fig.~\ref{fig3}(a)), which was
deduced by broadening the experimental curve (solid line) in order to smooth the gap-related structure. The method used
to obtain the normalized tunneling conductance is described in Ref.[\onlinecite{Vedeneev94}]. The energy gap, defined
as half of the peak-to-peak separation of the two main maxima in the $dI/dV(V)$ curves in Fig.~\ref{fig3}(b), is
$8$~meV at $1.8$~K.

The tunneling spectra presented in Fig.~\ref{fig3}(b) reveal a very large zero-bias conductance with only near $1.5\%$
conductance variation of the gap structures. We have not observed anything similar in the tunneling experiments with
Bi2212 single crystals in the superconducting state. In spite of numerous attempts, we were unable to obtain $dI/dV(V)$
curves with a sharp gap structure and a small zero-bias conductance as found for the superconducting break junctions on
the Bi2212.~\cite{Vedeneev05}

Figure~\ref{fig4}(a) displays the tunneling conductances $dI/dV(V)$ for the same tunnel break junction at $T=1.8$~K for
various magnetic fields applied along the c-axis of the crystal. For clarity, the curves have been shifted vertically
relatively the lowest curve. It is significant that there are no changes in the $dI/dV(V)$ curves with magnetic field,
notably the gap like structure does not change either in shape or in the voltage position (dashed vertical line).

Figure~\ref{fig4}(b) shows the tunneling conductances $dI/dV(V)$ for the same tunnel break junction in zero magnetic
field at various temperatures. The curves have been shifted vertically with respect to the upper one. As temperature
increases the gap like structure broadens and diminishes in amplitude rapidly with a little shift toward higher
voltages. Similar behavior of the tunneling spectra has been observed in studies of HTS at temperatures above $T_c$ by
using a scanning tunneling microscope (see \emph{e.g.} Ref. [\onlinecite{Renner98}]) and mesas, (see \emph{e.g.} Ref.
[\onlinecite{Suzuki99}]) providing evidence for the evolution of a superconducting gap in a pseudogap in the normal
state.

\begin{figure}
\includegraphics[width=8.5cm,angle=0,clip]{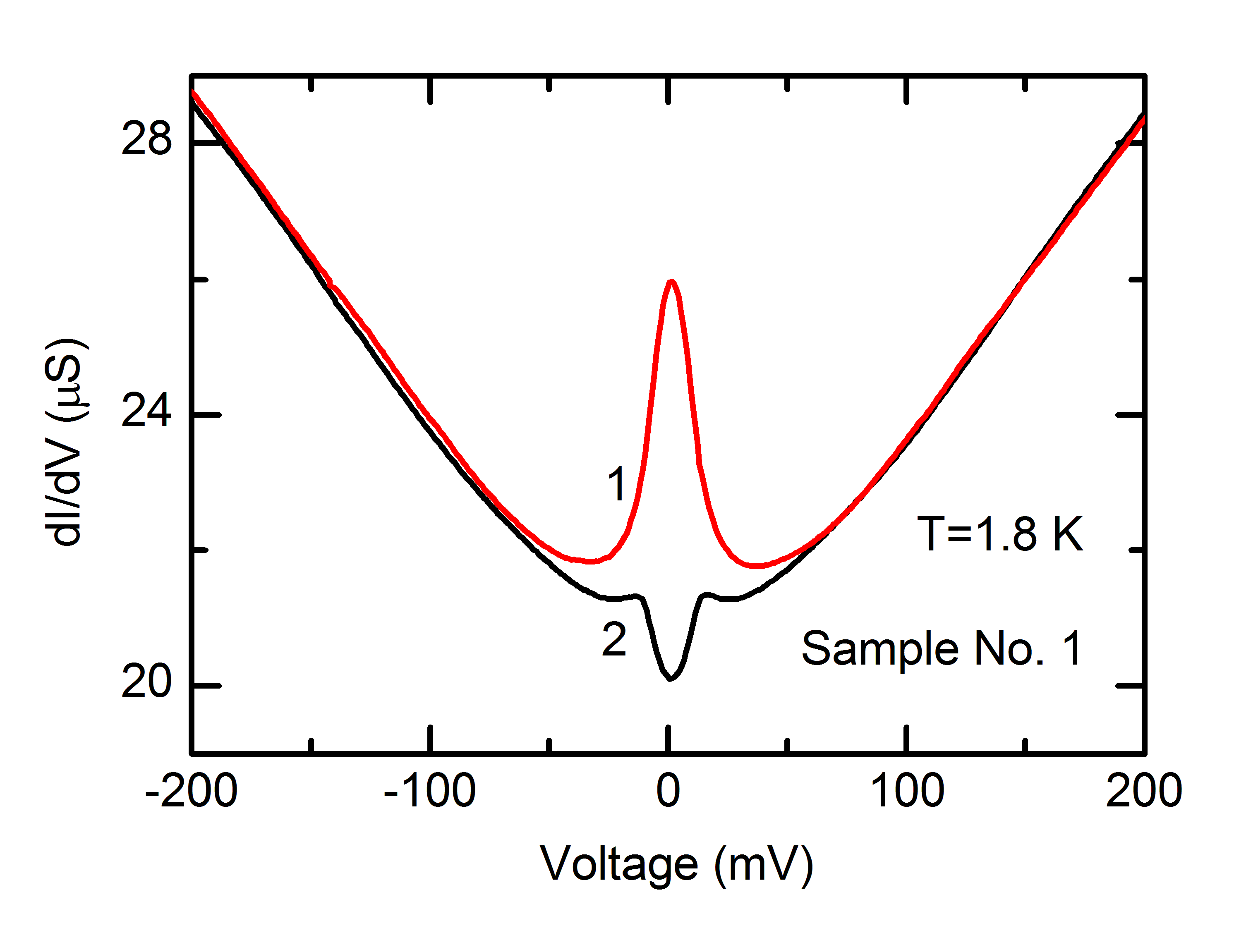}
\caption{\label{fig5} (color online) Zero-field tunneling conductances $dI/dV(V)$ for tunnel break junctions formed
along two different directions in the $ab$-plane at $T=1.8$~K. (Sample No. 1)}
\end{figure}

In our case it is simply not possible to relate the observed gap like structure in our $dI/dV$ spectra to the
superconducting gap: Firstly, because any ``tails'' of superconductivity in sample No. 1 are lacking (Fig.~\ref{fig1}),
Secondly, the voltage position of the gap peak in $dI/dV(V)$ remained unchanged in the applied magnetic fields
(Fig.~\ref{fig4}(a)), whereas, the superconducting gap peak simply must shift to lower voltages with increasing
field.~\cite{Vedeneev10} Thirdly, the superconducting gap peak must shift to lower voltages with increasing
temperature,~\cite{Vedeneev94} which emphatically is not what is observed in Fig.~\ref{fig4}(b). Thus it seems
reasonable to assume that the  gap like structure observed here in the $dI/dV$ spectra is associated with a some
``pseudogap'' in the electronic density of states, which exists in our nonsuperconducting cuprate. In this case such a
gap is either unrelated to superconductivity at all or the existence of such a gap is a necessary condition for the
subsequent occurrence of superconductivity with increasing carrier density in the sample.

Recent ARPES studies (see, \emph{e.g.} Lee \emph{et al.},\cite{Lee07} Kanigel \emph{et al.},\cite{Kanigel07} Chien
\emph{et al.}\cite{Chien09}) of Bi2212 showed that the pseudogap as well as the superconducting gap are anisotropic.
Above $T_c$, there is a gapless Fermi arc near the nodal region of momentum space and away from this Fermi arc region,
the well-known pseudogap gradually takes over and reaches its maximum at the antinodal region. In our break junctions
fabricated on single crystals of the layered cuprate superconductors, the tunneling occurs along the $ab$ plane. Since
the in-plane energy gaps are highly anisotropic, the tunneling conductance spectra can be quite different depending on
the tunneling direction relative to the crystalline axes, and do not always show maximum value of gap corresponding the
bulk density of states (see \emph{e.g.} Kashiwaya \emph{et al.},\cite{Kashiwaya95} Tanaka \emph{et
al.},\cite{Tanaka95}).

According to these articles, in the case of $ab$-plane tunneling, the main effect of varying the tunneling direction in
$d$-wave superconductors resides in the appearance of a zero-bias conductance peak for certain tunneling directions.
Needless to say, we could not reconstruct the break junction in such a way as to investigate all points of a Fermi
surface. Nevertheless, a number of the tunnel break junctions exhibited $dI/dV(V)$ curves with a zero-bias conductance
peak. Figure~\ref{fig5} displays the zero-field tunneling conductances $dI/dV(V)$ for tunnel break junctions which
appear to be along two distinctly different directions of the $ab$-plane for the same single crystal at $T=1.8$~K.
Curve 1 has the gap like structure as in Fig.~\ref{fig3} and Fig.~\ref{fig4}, but with a gap value equal to $15$~meV,
whereas, curve 2 has a zero-bias conductance peak. The large value of the peak relative to the gap structure in the
curve 1 is probably associated with different barrier heights. \cite{Kashiwaya95}

In our opinion, it is not surprising that the magnitude of the ``pseudogap'' observed here is much lower than the
typical values which are usually identified as the pseudogap. Similarly small values of the pseudogap have been
reported in underdoped Bi(La)2201,\cite{Kurosawa10} and in electron doped cuprates PrCeCuO.~\cite{Dagan05,Biswas01}
Furthermore, Kurosawa \textit{et al.}\cite{Kurosawa10} based on studies of the underdoped Bi(La)2201 using scanning
tunneling microscopy/spectroscopy showed that on the antinodal parts of the Fermi surface in momentum space there are
two different types of gap. One is the usual large pseudogap, whose gap width is much larger than that of a $d$-wave
pairing gap, and the other is the small pseudogap, whose gap width is comparable to that of simple $d$-wave gap. This
small pseudogap can open inside the large pseudogap at a lower temperature $T_c$.~\cite{Kurosawa10}

It is also important to note that the temperature and magnetic field dependence of the resistance and tunneling
measurements were performed on several nonsuperconducting single crystals similar to sample No.~1 and the extracted
magnitudes of the energy gap were in close agreement. However, obtaining a complete data set on a given single crystal
is difficult goal since tunnel break junctions are often unstable with time and changing temperature and magnetic
field. We show here measurements only for the single crystal for which we have a complete data set (temperature and
magnetic field dependence of the resistance and tunneling characteristic). However, we stress similar results have been
obtained on a number of other single crystals.

\section{Conclusion}

We have studied the magnetic field and temperature dependence of the in-plane tunneling conductance in high-quality non
superconducting (down to $10$~mK) layered oxycarbonate Bi$_{2+x}$Sr$_{4-x}$Cu$_2$CO$_3$O$_{8+\delta}$ single crystals.
Combining measurements of the in-plane magnetoresistivity $\rho_{ab}(T,H)$ and the magnetotunneling, we present
evidence for the existence of a small "pseudogap" in the nonsuperconducting cuprate without local incoherent pairs or
any correlation phenomena associated with superconductivity. This suggests that the observed "pseudogap" in HTS
cuprates is either unrelated to superconductivity at all, or that the existence of this gap may be a necessary
condition for the occurrence of superconductivity with increasing carrier density in the sample.

This work was partially supported by EuroMagNET II under the EU contract n. 228043.

%

\end{document}